\documentclass[12pt,thmsa]{article}
\usepackage{amsfonts}
\usepackage{graphicx}
\usepackage{epsfig}
\usepackage{graphics}

\makeatletter
\newcommand{\row}[1]{\mathord{\buildrel{\lower3pt\hbox{$\scriptscriptstyle\rightarrow$}}\over #1}}

\newcommand{\dyadic}[1]{\mathord{\dyadic@rrow{#1}}}
\newcommand{\dyadic@rrow}[1]{
\begin{picture}(12,12)(-1,0)
\put(-3,12){\makebox(0,0)[t]{$\scriptscriptstyle\downarrow$}}
\put(-3,13){\makebox(0,0)[l]{$\scriptscriptstyle\longrightarrow$}}
\put(5,0){\makebox(0,0)[b]{$#1$}}
\end{picture}
}

\input{tcilatex}

\begin{document}

\date{\today}

\begin{center}
{\large The quantum computational speed of a single Cooper Pair box}\\[0pt]
\vspace{0.5cm}  A.-S.F.Obada$^{1}$, D.A.M.Abo-Kahla$^{2}$, N. Metwally$^3$
and M. Abdel-Aty$^{3}$\\[0pt]
$^{1}$ Math. Dept., Faculty of Science, Al-Azhar University, Egypt \\[0pt]
$^{2}$Math. Dept., Faculty of Education,Ain Shams University, Egypt \\[0pt]
$^{3}$ Math. Dept., Faculty of Science, University of Bahrain, Bahrain
\end{center}

\begin{abstract}
The speed of computations is investigated by means of the orthogonality
speed for a charged qubit interacting with a single cavity field prepared
initially in a Fock state or Binomial state. We observe that the rate of the
computational speed is related to the number of photons inside the cavity.
Moreover, we show that the qubit-field coupling plays an opposite role,
where the speed of computations is decreased as the coupling is increased.
We suggest using the number of photons in the field as a control parameter
to improve the speed of computations.
\end{abstract}

pacs{74.70.-b, 03.65.Ta, 03.65.Yz, 03.67.-a, 42.50.-p} 

\section{Introduction}

In the last decade remarkable experiments were performed involving
measurements and manipulations of states for a single or several nanoscopic
Josephson junctions which were consistently interpreted in terms of
two-level quantum systems [1-5]. Low-capacitance Josephson-junction devices
have recently attracted a wide interest, both theoretically and
experimentally, particularly in view of the possibility of identifying
macroscopic quantum phenomena in their behavior. In this respect, one of the
circuits that have gained great attention is the so-called Superconducting
Cooper Pair Box (SCB), with an increasing number of experiments aimed at
supporting a qubit interpretation of its evolution [6-9]. Several schemes
have been proposed for implementing quantum computer hardware in solid state
quantum electronics. These schemes use electric charge [10], magnetic flux
[11] and superconducting phase [12] and electron spin [13].

The basic element of the quantum information is the quantum bit (qubit)
which is considered as a two level system. In the quantum information and
more precisely in the quantum computer, there is an important question which
would be raised: what is the speed of sending information from a nod to
another so as to reach the final output? Since the information is coded in a
density operator, we therefore ask how fast the density operator will change
its orthogonality [14]. This is to shed some light on the general behavior
of the interaction process and its relationship with the speed of the
computation [15-17] (maximum number of orthogonal states that the system can
pass through per unit time), speed of orthogonality [18] (minimum time for a
quantum state $\left\vert \Psi _{i}\right\rangle $ to evolve into orthogonal
state $\left\vert \Psi _{f}\right\rangle $ where $\langle \Psi
_{i}\left\vert \Psi _{f}\right\rangle =0$).

In the present paper, we consider the concrete situation of a two-level
system (Cooper pair box) interacting with a quantum cavity field. We
investigate the speed of computations when the initial state of the field is
considered either in a Fock state or a binomial state. The following
questions are considered: do useful properties of computational speed arise
from considering different initial state settings? what role, if any, does
mean photon number play any important role in the general behavior of the
computational speed? and, in the framework of an initial binomial state, is
it possible to obtain different orthogonality times which are useful for
quantum computation?  Answering these questions is the main aim of this
paper.

The paper is organized as follows: in section II, we introduce a brief
discussion on the qubit-field interaction and its dynamics. Section III is
devoted to discuss the measures of the speed of the computations of typical
bipartite states. Finally, discussion of the results and conclusion are
given.

\begin{figure}[t]
\begin{center}
\includegraphics[width=15cm,height=6cm]{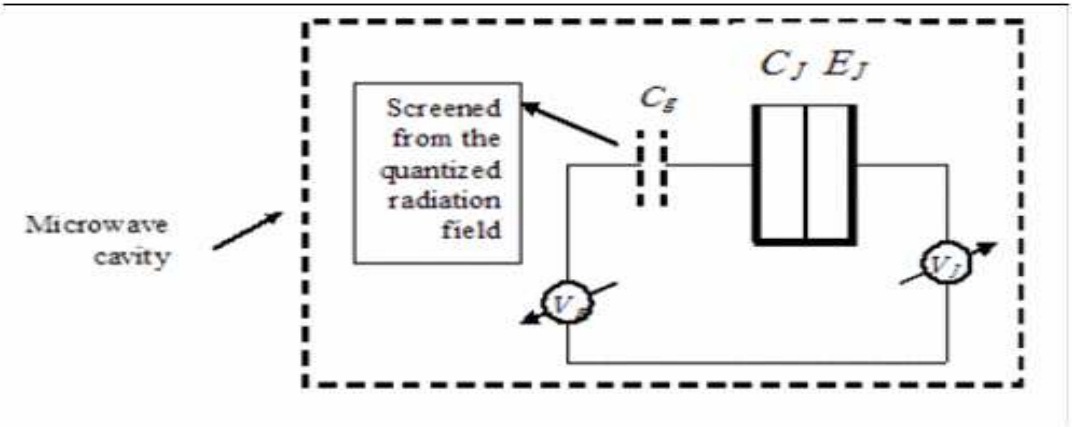}
\end{center}
\caption{Schematic picture of the Cooper-pair box which is driven by an
applied voltage, $Vg,$ through the gate capacitance, $Cg$. Black bars denote
Cooper-pair box. The two Josephson junctions have capacitance, $C_{J},$ and
Josephson energy, $E_{J}.$ The driving microwave field is generated using
the electrical voltage acting on the charge qubit via the gate capacitance
[11].}
\end{figure}

\section{The model}

We consider a superconducting box connected by a low-capacitance Josephson
junction with capacitance $C_{J}$ and Josephson energy $E_{J}$, coupled
capacitively to a gate voltage $V_{g}$ (gate capacitance $C_{g}$), placed
inside a single-mode microwave cavity. We suppose that the gate capacitance $%
C_{g}$ is screened from the quantized radiation field (see figure (1)), and
then the Hamiltonian of the system can be written as [19-21]

\begin{equation}
H=\frac{(Q-C_{g}V_{g}-C_{J}V)^{2}}{2(C_{g}+C_{J})}-E_{J}\cos \phi +\hbar
\omega (a^{\dagger }a+\frac{1}{2}),
\end{equation}%
where $Q=2Ne$\ is the Cooper pair charge on the island, where $N$ is the
number of Cooper-pairs, $\phi $\ is the phase difference across the
junction, $\omega $\ is the field frequency, and $a^{+}$, $a$ are the
creation and annihilation operators of the microwave. $V$ is the effective
voltage difference produced by the microwave across the junction. $V$\ may
be written down as [19-21]

\begin{equation}
V=i(\frac{\hbar \omega }{2C_{F}})^{\frac{1}{2}}(a-a^{\dagger }),
\end{equation}%
where $C_{F}$ is the capacitance parameter, which depends on the thickness
of the junction, the relative dielectric constant of the thin insulating
barrier, and the dimension of the cavity. Here, we consider the case where
the charging energy with scale $E_{c}=\frac{e^{2}}{2(C_{g}+C_{J})}$ dominate
over the Josephson coupling energy $E_{J}$ , and concentrate on the value $%
V_{g}=\frac{e}{C_{g}}$ and weak quantized radiation field, so that only the
two low-energy charge states $N=0$ and $N=1$ are relevant. In this case the
Hamiltonian in a basis of the charge state $\left\vert \downarrow
\right\rangle $ and $\left\vert \uparrow \right\rangle $ reduces to a
two-state form in a spin- $1/2$ language [10,23-24]%
\begin{equation}
H=E_{c}(1+\frac{C_{J}^{2}V^{2}}{e^{2}})-2E_{c}\frac{C_{J}V}{e}J_{z}-\frac{1}{%
2}E_{J}J_{x}+\hbar \omega (a^{\dagger }a+\frac{1}{2}).
\end{equation}%
We denote by $J_{z}$ and $J_{x}$ the Pauli matrices in the pseudo-spin basis
\{$\left\vert \downarrow \right\rangle $, $\left\vert \uparrow \right\rangle
$\},%
\begin{equation}
J_{x}=\left\vert \uparrow \right\rangle \left\langle \downarrow \right\vert
+\left\vert \downarrow \right\rangle \left\langle \uparrow \right\vert %
\mbox{ and }J_{z}=\left\vert \uparrow \right\rangle \left\langle \uparrow
\right\vert -\left\vert \downarrow \right\rangle \left\langle \downarrow
\right\vert \mbox{},
\end{equation}%
where the charge states are not the eigenstates of the Hamiltonian (3), even
in the absence of the quantized radiation field, i.e.$V=0$, we describe $H$
in the two charge states subspace through new states and denote the
corresponding states as $\left\vert +\right\rangle $\ and $\left\vert
-\right\rangle $\ [24] as%
\begin{equation}
\left\vert +\right\rangle =\frac{1}{\sqrt{2}}(\left\vert \uparrow
\right\rangle -\left\vert \downarrow \right\rangle )\mbox{,
}\left\vert -\right\rangle =\frac{1}{\sqrt{2}}(\left\vert \uparrow
\right\rangle +\left\vert \downarrow \right\rangle ).
\end{equation}%
In the weak quantized radiation field, one may neglect the term containing $%
V^{2}$ in equation (3) and from Eqs. (1-5), the Hamiltonian in the new basis
$\left\vert +\right\rangle $ and $\left\vert -\right\rangle $ is given by

\begin{equation}
H=E_{c}-2iE_{c}\frac{C_{J}}{e}(\frac{\hbar \omega }{2C_{F}})^{\frac{1}{2}%
}(a-a^{\dagger })\sigma _{x}+\frac{1}{2}E_{J}\sigma _{z}+\hbar \omega
(a^{\dagger }a+\frac{1}{2}).
\end{equation}%
We denote by $\sigma _{z}$ the Pauli matrix, $\sigma _{+}$ and $\sigma _{-}$
the raising and lowering operators ($\left[ \sigma _{+},\sigma _{-}\right]
=\sigma _{z}$). In the rotating wave approximation,the Hamiltonian takes the
following form

\begin{equation}
H=E_{c}+\frac{1}{2}E_{J}\sigma _{z}+i\hbar g(a\mbox{ }\sigma
_{+}-a^{\dagger }\sigma _{-})+\hbar \omega (a^{\dagger
}a+\frac{1}{2})\mbox{,},
\end{equation}%
where
\begin{eqnarray}
g &=&(\frac{\omega }{C_{F}\hbar })^{\frac{1}{2}}\frac{E_{c}}{2} \\
&=&(\frac{\omega }{2C_{F}\hbar })^{\frac{1}{2}}\frac{eC_{J}}{(C_{g}+C_{J})}%
\mbox{.}.
\end{eqnarray}%
It is noted that the Hamiltonian (7) is just like the simplest form of atom-
field interaction, which is known as the Jaynes-Cummings model (JCM) [25].
In this paper we consider the case where $E_{J}\sim \hbar \omega <<E_{c}$
then, in the interaction picture, the Hamiltonian (7) takes the form ($\hbar
=1$),%
\begin{equation}
H_{I}^{I}=\frac{1}{2}\Delta \sigma _{z}+ig(a\mbox{ }\sigma _{+}-a^{\dagger
}\sigma _{-}),
\end{equation}%
where $\Delta =E_{J}-\omega $ is the detuning between the Josephson energy
and cavity field frequency. We shall be working from now on in the basis $%
\left\{ \left\vert +\right\rangle ,\left\vert -\right\rangle \right\} $,
then in the interaction picture, the Hamiltonian (9) is written as [26]%
\begin{equation}
H_{I}^{I}=\left(
\begin{array}{cc}
\frac{\Delta }{2} & -iga \\
iga^{\dagger } & -\frac{\Delta }{2}%
\end{array}%
\right) ,
\end{equation}%
and the corresponding evolution operator $U_{t}=\exp (-iH_{I}^{I}t)$ can be
written in the form%
\begin{equation}
U_{t}=\left(
\begin{array}{cc}
U_{11}(t) & U_{12}(t) \\
U_{21}(t) & U_{22}(t)%
\end{array}%
\right) ,
\end{equation}%
where%
\begin{eqnarray}
U_{11}(t) &=&\cos \Omega _{n+1}t-i\frac{\Delta }{2}\frac{\sin \Omega _{n+1}t%
}{\Omega _{n+1}},  \nonumber  \label{Unitary} \\
U_{12}(t) &=&iga\frac{\sin \Omega _{n}t}{\Omega _{n}},  \nonumber \\
U_{22}(t) &=&\cos \Omega _{n}t+i\frac{\Delta }{2}\frac{\sin \Omega _{n}t}{%
\Omega _{n}},
\end{eqnarray}%
with, $\Omega _{n}=(\frac{\Delta ^{2}}{4}+g^{2}n)^{2},n=a^{\dagger }a$ and $%
U_{21}=U_{12}^{\dagger }$. The density operator at any time, $t>0,$ is given
by
\begin{equation}
\rho _{I}(t)=U_{t}\rho (0)U_{t}^{\dagger },
\end{equation}%
where$\rho (0)=\rho _{b}(0)\otimes \rho _{f}(0)$. Having obtained the
density operator at any time, $t\geq 0$ ,one can investigate the speed of
computation as described in the following sections.

\section{The speed of computation}

We shall assume that the box is initially in its pure state, that is

\begin{equation}
\rho _{b}(0)=\frac{1}{2}\left(
\begin{array}{cc}
1 & 1 \\
1 & 1%
\end{array}%
\right) ,
\end{equation}%
with eigenvectors

\begin{equation}
\nu _{\pm }=\frac{1}{\sqrt{2}}\left(
\begin{array}{c}
1 \\
\pm 1%
\end{array}%
\right) .
\end{equation}%
Using this initial state, the density operator at any time, $t>0,$\ is
given\ by%
\begin{equation}
\rho _{I}(t)=\left(
\begin{array}{cc}
\rho _{11}(t) & \rho _{12}(t) \\
\rho _{21}(t) & \rho _{22}(t)%
\end{array}%
\right) ,
\end{equation}

where,%
\begin{eqnarray}
\rho _{11}(t) &=&\frac{1}{2}(U_{11}\rho _{f}(0)U_{11}^{\dagger }+U_{12}\rho
_{f}(0)U_{11}^{\dagger }+U_{11}\rho _{f}(0)U_{12}^{\dagger }+U_{12}\rho
_{f}(0)U_{12}^{\dagger }),  \nonumber \\
\rho _{12}(t) &=&\frac{1}{2}(U_{11}\rho _{f}(0)U_{21}^{\dagger }+U_{12}\rho
_{f}(0)U_{21}^{\dagger }+U_{11}\rho _{f}(0)U_{22}^{\dagger }+U_{12}\rho
_{f}(0)U_{22}^{\dagger }),  \nonumber \\
\rho _{22}(t) &=&\frac{1}{2}(U_{21}\rho _{f}(0)U_{21}^{\dagger }+U_{22}\rho
_{f}(0)U_{21}^{\dagger }+U_{21}\rho _{f}(0)U_{22}^{\dagger }+U_{22}\rho
_{f}(0)U_{22}^{\dagger }),
\end{eqnarray}

and $\rho_{21}(t)=\rho^{*}_{12}(t)$.

\subsection{Initial Fock State}

Now we assume that the field is initially in a Fock state, that is $\rho
_{f}(0)=\left\vert n\right\rangle \left\langle n\right\vert $, then one can
calculate $\rho _{ij}^{b}(t)$ by calculating $Tr_{f}(U\rho _{f}(0)U^{\dagger
})_{ij}$, The elements of the density operator are given by,%
\begin{eqnarray}
\rho^{b} _{11}(t) &=&\frac{1}{2}\Bigl(\cos ^{2}\Omega _{n+1}t+(\frac{\Delta
}{2})^{2}\frac{\sin ^{2}\Omega _{n+1}t}{\Omega _{n+1}^{2}}+g^{2}n\frac{\sin
^{2}\Omega _{n}t}{\Omega _{n}^{2}}\Bigr),  \nonumber \\
\rho^{b} _{22}(t) &=&\frac{1}{2}\Bigl(\cos ^{2}\Omega _{n}t+(\frac{\Delta }{2%
})^{2}\frac{\sin ^{2}\Omega _{n}t}{\Omega _{n}^{2}}+g^{2}(n+1)\frac{\sin
^{2}\Omega _{n+1}t}{\Omega _{n+1}^{2}}\Bigr),  \nonumber \\
\rho^{b} _{12}(t) &=&\frac{1}{2}\Bigl(\cos \Omega _{n+1}t-i\frac{\Delta }{2}%
\frac{\sin \Omega _{n+1}t}{\Omega _{n+1}})(\cos \Omega _{n}t-i\frac{\Delta }{%
2}\frac{\sin \Omega _{n}t}{\Omega _{n}}\Bigr).
\end{eqnarray}
\begin{figure}[tbp]
\begin{center}
\includegraphics[width=18pc,height=12pc]{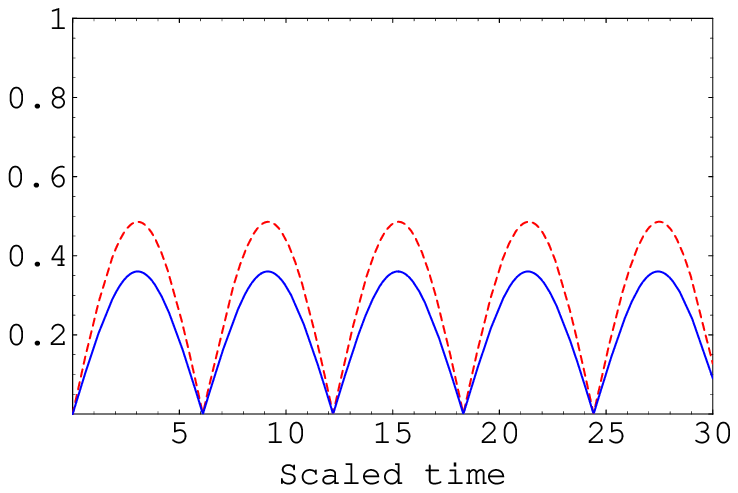} %
\includegraphics[width=18pc,height=12pc]{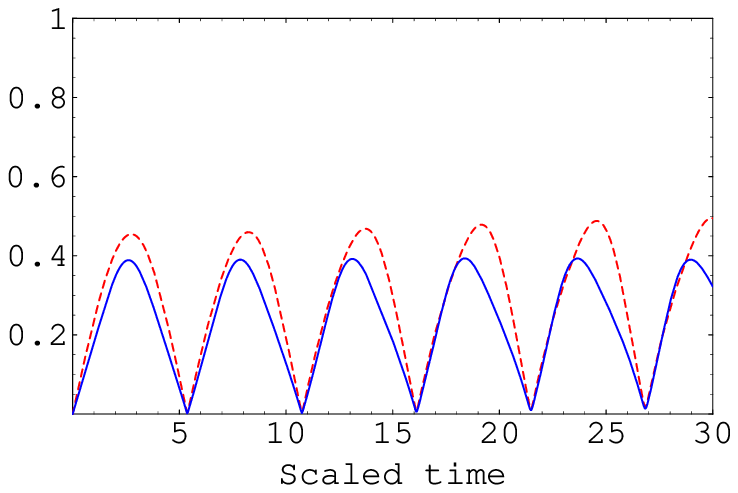} \put(-250,130){a}
\put(-25,130){b} \\[0pt]
\includegraphics[width=18pc,height=12pc]{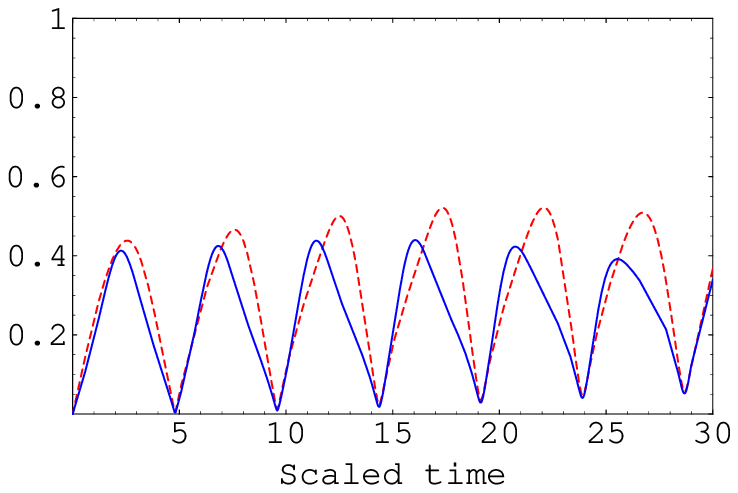} %
\includegraphics[width=18pc,height=12pc]{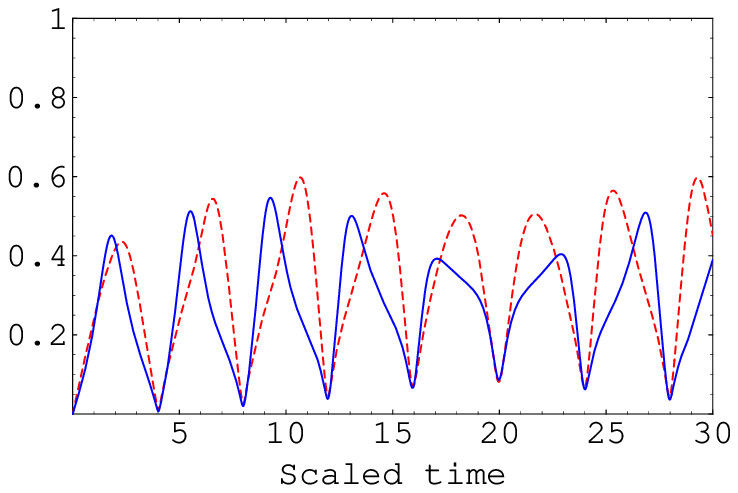} \put(-250,130){c}
\put(-25,130){d}
\end{center}
\caption{The speed of orthogonality for $\Delta=n=1$ (a) $g=0.1$ and (b) $%
g=0.25$, (c) $g=0.35$ and (d)$g=0.5$.}
\end{figure}
Direct calculations can be used to obtain the eigenvectors $u_{1,2}(t)$\ for
the final state $\rho^{b}(t),$ we can explicitly write $u_{1,2}(t)$\ as the
follows%
\begin{equation}
u_{1,2}(t)=\pm \sqrt{\frac{\left\vert \rho^{b} _{12}(t)\right\vert ^{2}}{%
\left\vert \rho^{b} _{12}(t)\right\vert ^{2}+\left\vert \lambda
_{1,2}(t)-\rho^{b} _{11}(t)\right\vert ^{2}}}\left(
\begin{array}{c}
1 \\
\frac{(\lambda _{1,2}(t)-\rho^{b} _{11}(t))}{\rho^{b} _{12}(t)}%
\end{array}%
\right) ,
\end{equation}%
where%
\begin{equation}
\lambda _{1,2}(t)=\frac{1}{2}\left( 1\pm \sqrt{(\rho^{b}
_{11}(t)-\rho^{b}_{22}(t))^{2}+4\left\vert \rho^{b} _{12}(t)\right\vert ^{2}}%
\right) .
\end{equation}%
In order to facilitate our discussion, let us define the scalar product of
the vectors $u_{i}(t)$ and $\nu _{j}(0)$\ such as
\begin{equation}
Sp_{ij}=\langle \nu _{i}(0)\left\vert u_{j}(t)\right\rangle,
\end{equation}%
where $\nu_i$ and $u_j$ represent the eigenvectors of the initial and final
state of the Cooper-pair box. The expression $Sp_{ij}$ represents the dot
product of the four possibilities i.e $Sp_{11}= \langle \nu
_{1}(0)\left\vert u_{1}(t)\right\rangle$, $Sp_{12}= \langle \nu
_{1}(0)\left\vert u_{2}(t)\right\rangle$. If the dot product for any two
eigenvectors vanishes this means that the two vectors are orthogonal and the
information which is coded in one eigenvector is transformed to the other
eigenvector. The number of vanishing eigenvectors indicates speed of
orthogonality and consequently the speed of computations.

It should be noted that in our calculations we have taken into account all
the possible products of $u_{i}$\ and $\nu _{j}$, but we produce the best
figures in all the possible products of $u_{i}$\ and $\nu _{j}$. In what
follows we present the dynamics of the amplitude values of $Sp_{ij}$ which
represents the speed of orthogonality against the scaled time for different
values of the field and the Cooper pair parameters.

In Fig. 2, we investigate the effect of the coupling constant $g$ on the
speed of orthogonality $Sp_{ij}$, where, we set $\Delta =n=1$. It is clear
that for small values of the coupling constant, $g$, the speed of
orthogonality is very large. As the coupling constant is increased the speed
of orthogonality is decreased. However for $g>0.5$, the speed is almost zero.

\begin{figure}[tbp]
\begin{center}
\includegraphics[width=18pc,height=12pc]{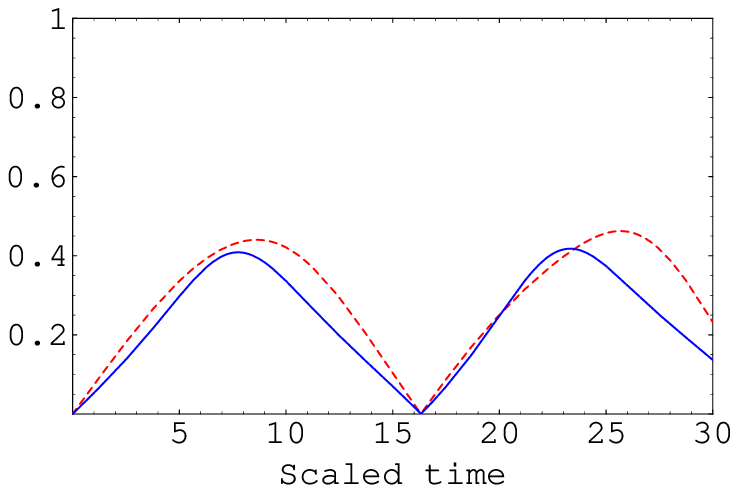} %
\includegraphics[width=18pc,height=12pc]{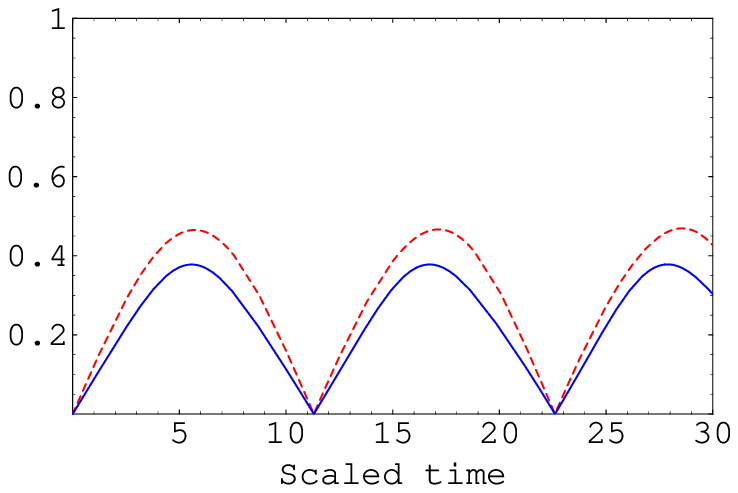} \put(-250,130){a}
\put(-25,130){b} \\[0pt]
\includegraphics[width=18pc,height=12pc]{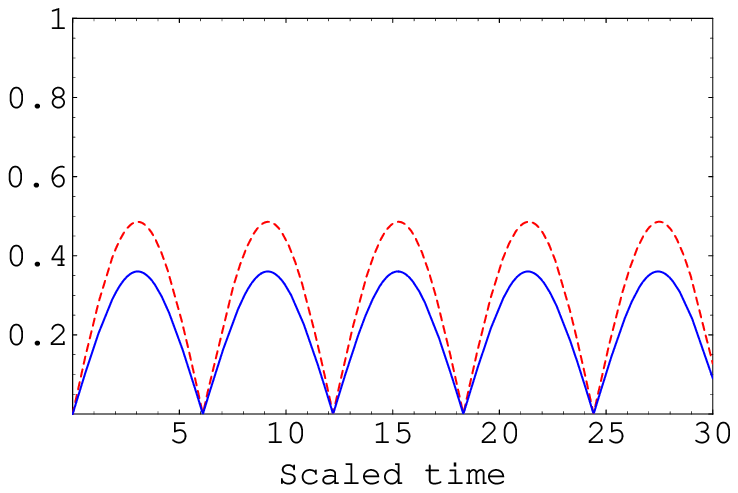} %
\includegraphics[width=18pc,height=12pc]{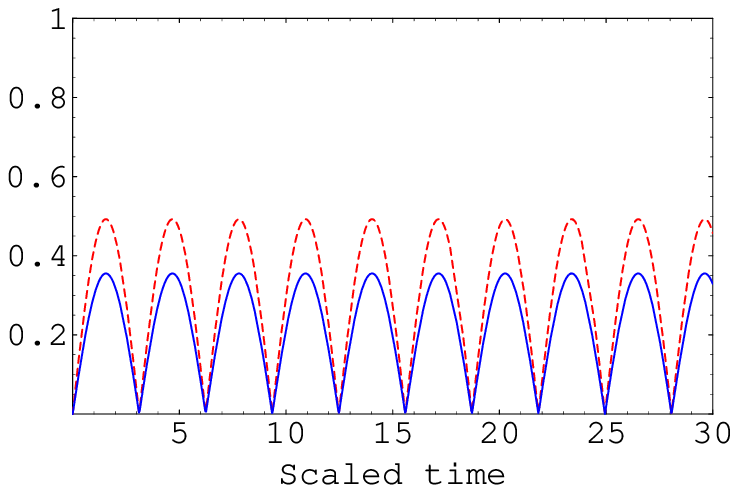} \put(-250,130){c}
\put(-25,130){d}
\end{center}
\caption{The effect of the detuning parameter on the speed of orthogonality.
The parameters are $n=1$, $g=0.1$ and different values of detuning, where,
(a)$\Delta =0.3$ (b) $\Delta =0.5$ and (c) $\Delta =1$, (d) $\Delta =2$.}
\end{figure}
\begin{figure}[tbp]
\begin{center}
\includegraphics[width=18pc,height=12pc]{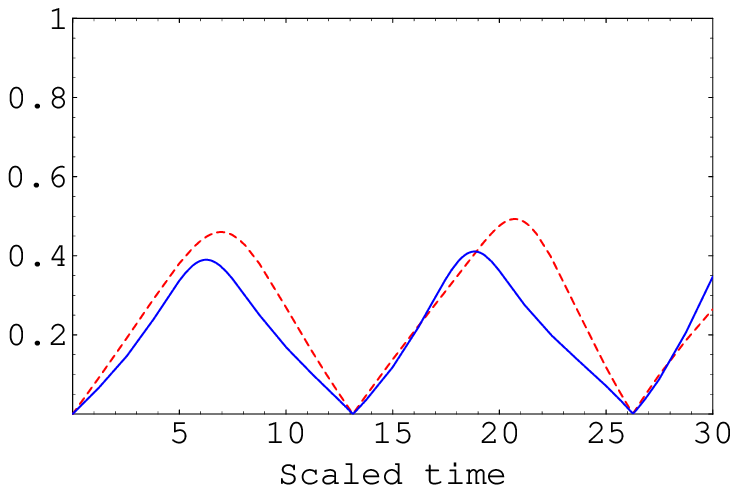} %
\includegraphics[width=18pc,height=12pc]{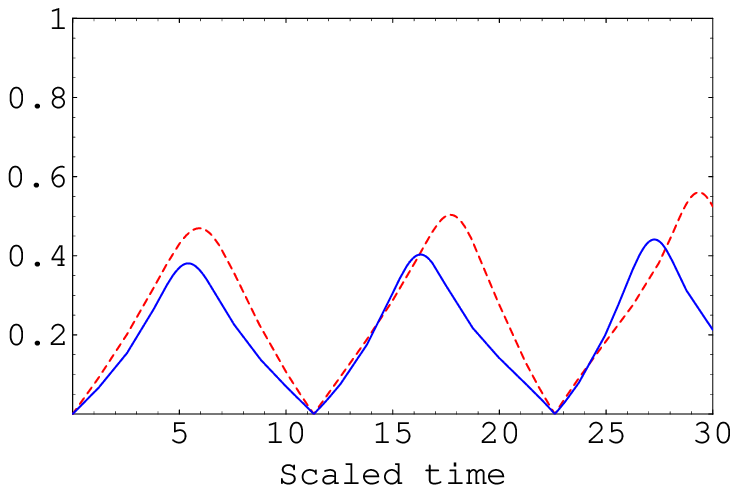} \put(-250,130){a}
\put(-25,130){b} \\[0pt]
\includegraphics[width=18pc,height=12pc]{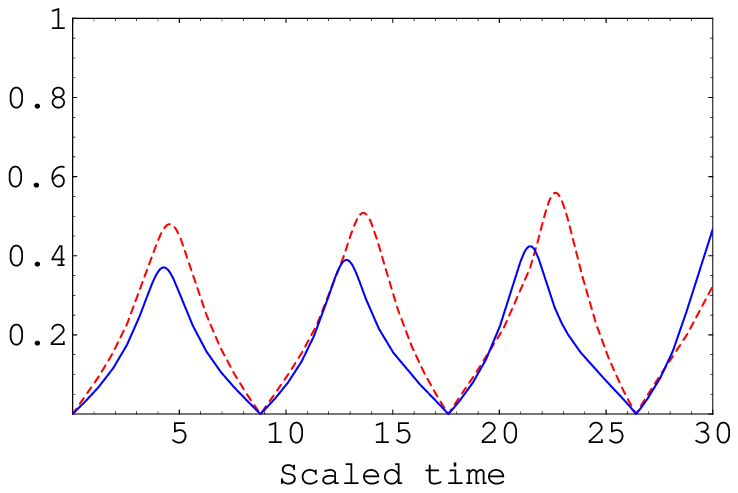} %
\includegraphics[width=18pc,height=12pc]{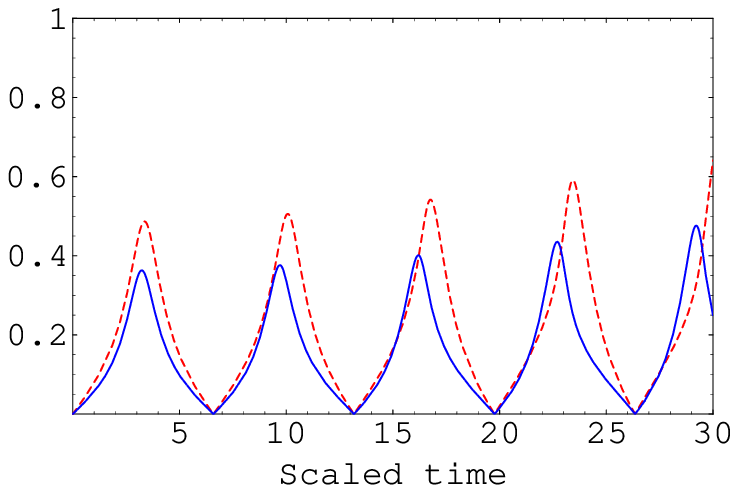} \put(-250,130){c}
\put(-25,130){d}
\end{center}
\caption{The effect of the number of photons on the speed of orthogonality,
where $g=0.1$, $\Delta =0.3$(a)$n=3$ (b) $n=5$ and (c) $n=10$, (d) $n=20$.}
\end{figure}

Fig. 3 shows the dynamics of orthogonality speed for different values of the
detuning parameter, where the other parameters are assumed to be constant.
In Fig. 3a we set a small value of the detuning parameter $\Delta =0.3$,
while the number of photons $n=1$ and the coupling constant $g=0.1$. It is
clear that  the orthogonality appears only one time. However as one
increases the detuning, $\Delta =0.5,$ as shown in Fig. 3b, the amplitude of
$Sp_{ij}$ vanishes at a specific time in this range of the scaled time. This
means that the speed of orthogonality is increased. Further increase of the
detuning parameter, the number of zeros of the amplitudes of $Sp_{ij}$ is
increased and consequently the speed of orthogonality is increased as shown
in Fig. 3c and Fig.3d, where we set $\Delta =1$ and $\Delta =2,$
respectively.

Now, let us investigate the effect of the field parameter which is
represented by the number of photons $n$ on the speed of orthogonality. In
Fig. 4, we consider different values of the photon number $n$, while we
consider the values of the other parameters such that the speed of
orthogonality is very small. Fig. 4a displays the dynamical behavior of the
amplitude $Sp_{ij}$ for $n=3$. One sees that, this amplitude vanishes once,
which means that the orthogonality appears another time (see Fig. (3a),
where $n=1$). This means that the speed of orthogonality is increased as one
increases the number of photon $n$. This remark appears clearly in Fig. 4b,
Fig. 4c and Fig. 4d, where we set $n=5,10$ and $20,$ respectively.

From Figs. 2 and 3, it is clear that as one increases the coupling constant,
$g$, i.e., the ratio $C_{j}/(C_{g}+C_{j})$ is increased, the possibility of
increasing the speed of orthogonality is lower. One can overcome this
problem by increasing the detuning parameter. However, it may be difficult
to control the parameters $g$ and $\Delta $, but it will be more easy to
control the number of photons $n$. In this case, by increasing the number of
photons one can increases the speed of orthogonality and consequently the
speed of computation. Therefore, one can look at these parameters as control
parameters to improve the speed of computations.

\subsection{Initial Binomial State}

\begin{figure}[tbp]
\begin{center}
\includegraphics[width=18pc,height=12pc]{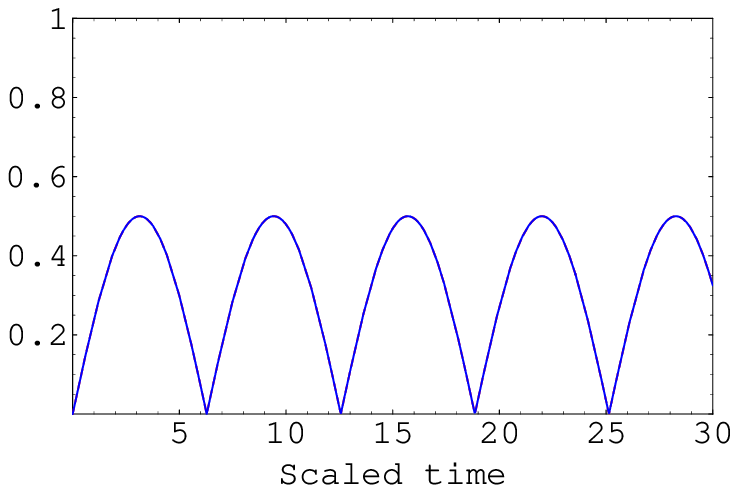} %
\includegraphics[width=18pc,height=12pc]{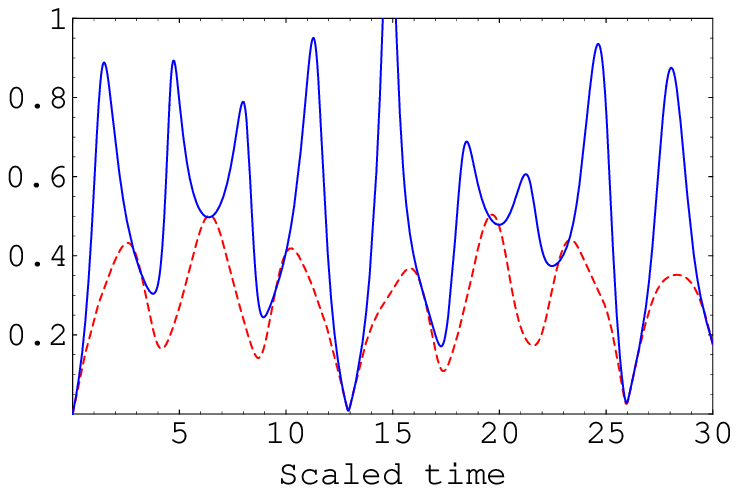} \put(-250,130){a}
\put(-25,130){b} \\[0pt]
\end{center}
\caption{The speed of orthogonality for a Cooper pair interact with a field
in prepared initially in a binomial state, where $\Delta =1$ and $\protect%
\eta =0.1$ (a) $g=0.001$ and (b) $g=0.8$.}
\end{figure}

In this case, we assume that the field is initially prepared in a binomial
state,

\begin{equation}
\rho _{f}(0)=\left\vert \mu ,\eta \right\rangle \left\langle \eta ,\mu
\right\vert  \label{FiledB}
\end{equation}%
where

\begin{equation}
\left\vert \mu ,\eta \right\rangle =\sum_{m=0}^{\mu }\sqrt{\kappa }\eta
^{m}(1-\left\vert \eta \right\vert ^{2})^{\frac{\mu -m}{2}}\left\vert
m\right\rangle ,\quad \kappa =\frac{\mu !}{m!(\mu -m)!},\mu \mbox{ an
integer, }\left\vert \eta \right\vert \leq 1,
\end{equation}

the coherent state is obtained when $(\mu \rightarrow \infty ,$and $%
\left\vert \eta \right\vert \rightarrow 0$ such that $\mu \eta
^{2}\rightarrow \tilde{n}=\left\vert \alpha \right\vert ^{2}).$The time
evaluation of the density operator of the charged qubit and the field, $\rho
(t),$ is obtained by using (\ref{FiledB}) and the unitary operator (\ref%
{Unitary}). Tracing out the field state, one obtains the time evolution of
the qubit state, namely, $\rho _{ij}^{b}(t)=Tr_{f}(U\rho _{f}(0)U^{\dagger
})_{ij}$. The elements of this density operator are given by,%
\begin{eqnarray}
\rho _{11}(t) &=&\sum_{n=0}^{\mu }\Upsilon\eta ^{2n}(1-\left\vert \eta
\right\vert ^{2})^{\mu -n}[\cos ^{2}\Omega _{n+1}t+g^{2}n\frac{\sin
^{2}\Omega _{n}t}{\Omega _{n}^{2}}  \nonumber \\
&&+(\frac{\Delta }{2})^{2}\frac{\sin ^{2}\Omega _{n+1}t}{\Omega _{n+1}^{2}}%
-2g\frac{(\mu -n)}{\sqrt{n+1}}\eta (1-\left\vert \eta \right\vert ^{2})^{-%
\frac{1}{2}}\cos \Omega _{n+1}t\frac{\sin \Omega _{n}t}{\Omega _{n}}],
\nonumber \\
\rho _{22}(t) &=&\sum_{n=0}^{\mu }\Upsilon\eta ^{2n}(1-\left\vert \eta
\right\vert ^{2})^{\mu -n}[\cos ^{2}\Omega _{n}t+g^{2}(n+1)\frac{\sin
^{2}\Omega _{n+1}t}{\Omega _{n+1}^{2}}  \nonumber \\
&&+(\frac{\Delta }{2})^{2}\frac{\sin ^{2}\Omega _{n}t}{\Omega _{n}^{2}}-2g%
\frac{(\mu -n)}{\sqrt{n+1}}\eta (1-\left\vert \eta \right\vert ^{2})^{-\frac{%
1}{2}}\cos \Omega _{n+1}t\frac{\sin \Omega _{n+1}t}{\Omega _{n+1}}],
\nonumber \\
&&  \nonumber \\
\rho _{12}(t) &=&\sum_{n=0}^{\mu }\Upsilon\eta ^{2n}(1-\left\vert \eta
\right\vert ^{2})^{\mu -n}[(\cos \Omega _{n+1}t-i\frac{\Delta }{2}\frac{\sin
\Omega _{n+1}t}{\Omega _{n+1}})(\cos \Omega _{n}t  \nonumber \\
&&-i\frac{\Delta }{2}\frac{\sin \Omega _{n}t}{\Omega _{n}})-g^{2}\frac{(\mu
-n)}{\sqrt{n+1}}\frac{(\mu -n+1)}{\sqrt{n+2}}(1-\left\vert \eta \right\vert
^{2})^{-1}\frac{\sin \Omega _{n+1}t}{\Omega _{n+1}}\frac{\sin \Omega _{n+2}t%
}{\Omega _{n+2}}  \nonumber \\
&&+g\frac{(\mu -n)}{\sqrt{n+1}}\eta (1-\left\vert \eta \right\vert ^{2})^{-%
\frac{1}{2}}\frac{\sin \Omega _{n+1}t}{\Omega _{n+1}}[(\cos \Omega
_{n}t-\cos \Omega _{n+2}t)  \nonumber \\
&&+i\frac{\Delta }{2}(\frac{\sin \Omega _{n}t}{\Omega _{n}}-\frac{\sin
\Omega _{n+2}t}{\Omega _{n+2}})]],
\end{eqnarray}%
where $\Upsilon=\frac{\mu!}{(\mu-n)! n!}$.  Using the obtained density
operator of the system, we calculate the speed of orthogonality and
consequently the computational speed.

The effect of the coupling constant on the speed of orthogonality for a
Cooper pair interacts with a cavity mode initially prepared in the binomial
state is described in Fig.5. It is clear that, as one increases the coupling
constant the speed of orthogonality decreases, where the numbers of
vanishing amplitudes $Sp_{ij}$ increases for larger values of $g$. However
for $g>1$, the amplitudes $Sp_{ij}$ oscillate very fast but the number of
orthogonality decreases.
\begin{figure}[tbp]
\begin{center}
\includegraphics[width=18pc,height=12pc]{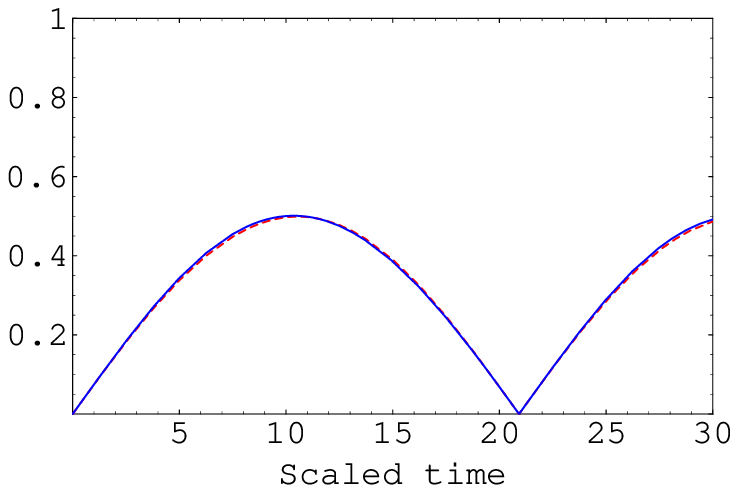} %
\includegraphics[width=18pc,height=12pc]{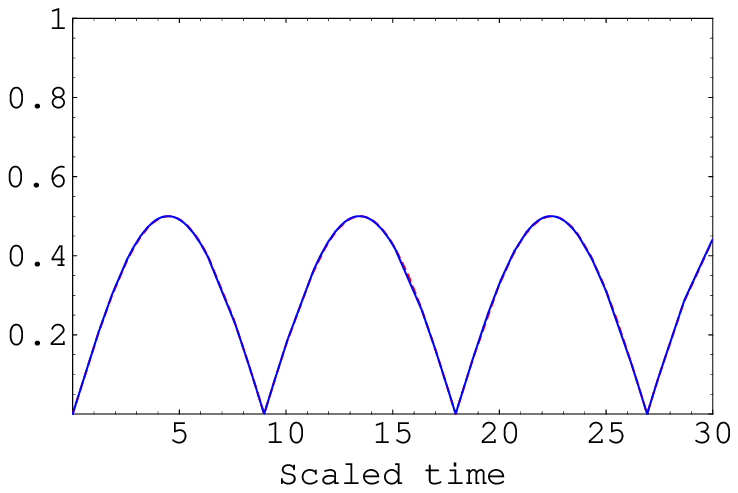} \put(-245,128){a}
\put(-25,130){b} \\[0pt]
\includegraphics[width=18pc,height=12pc]{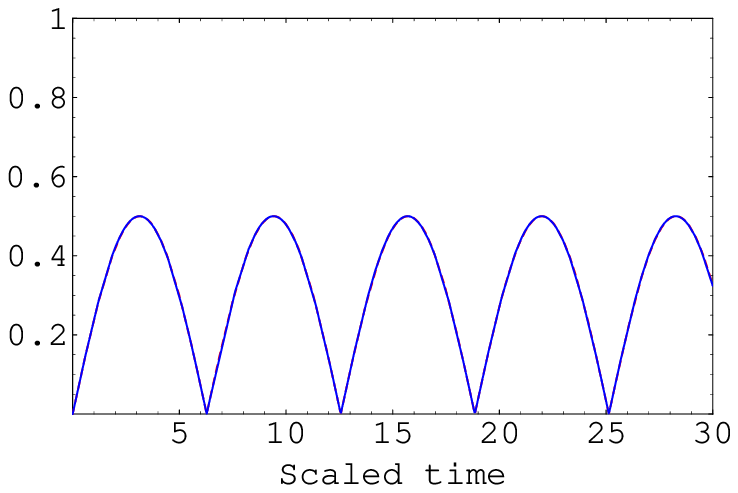} %
\includegraphics[width=18pc,height=12pc]{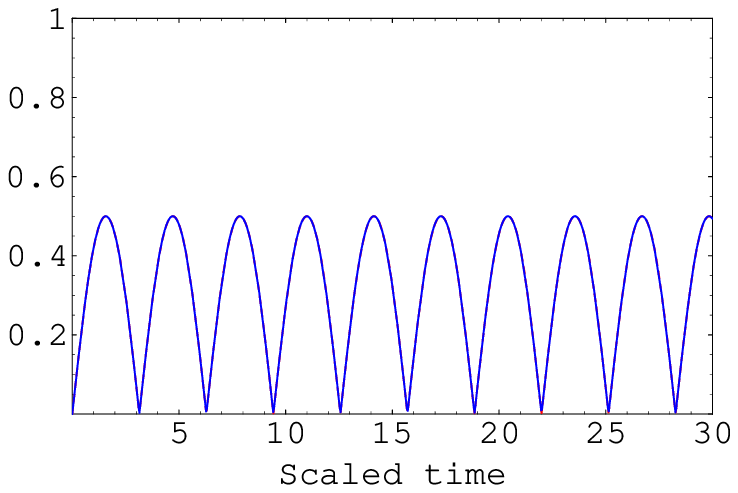} \put(-248,128){c}
\put(-25,130){d} \\[0pt]
\end{center}
\caption{The same as Fig.(4), where $g=0.01,\protect\eta=0.1$ and (a) $%
\Delta=0.3$ and (b) $\Delta=0.7$, (c) $\Delta=1$ and (d)$\Delta=2$.}
\end{figure}
\begin{figure}[tbp]
\begin{center}
\includegraphics[width=18pc,height=12pc]{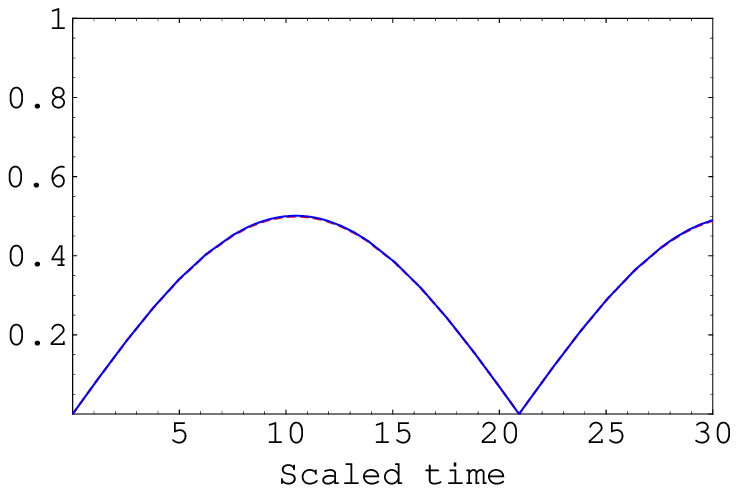} %
\includegraphics[width=18pc,height=12pc]{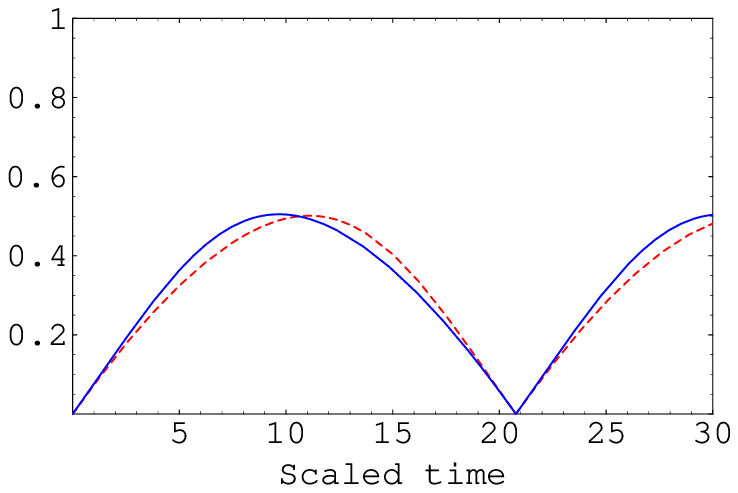} \put(-250,128){a}
\put(-30,130){b} \\[0pt]
\end{center}
\caption{The same as Fig.(4), where $g=0.01$, $\Delta =0.3$ and (a) $\protect%
\eta =0.001$ and (b) $\protect\eta =0.8$.}
\end{figure}

Fig.6 describe the dynamics of the amplitudes $Sp_{ij}$ for different values
of the detuning parameter while the other parameters are assumed to be
fixed. It is clear that, for small value of the detuning, $\Delta =0.3$ the
number of vanishing $Sp_{ij}$ is small and consequently the speed of
orthogonality. However as one increases the detuning the speed of
orthogonality increases and consequently the speed of computations. On the
other hand, if we compare Fig. 6a, where $\Delta =0.3$ and Fig. 6b, where $%
\Delta =0.5$, we can see that the amplitudes $Sp_{ij}$ vanishes two times
for the latter case. Moreover, the orthogonality time, the time in which the
amplitudes vanish, is shorter than that depicted in Fig. 6a. However, for
larger values of the detuning parameter the orthogonality time decreases
more and consequently the computational time, the time which is taken to
transfer the information from one node to another, decreases. Therefore, it
will be enough to investigate the effect of the parameter $\eta $ on the
orthogonality speed. Fig.7, displays the dynamics of $Sp_{ij}$ for different
values of the parameter $\eta $. In Fig. 7a, we set a small value of $\eta ,$
(say $\eta =0.001)$, while the other parameters are fixed. However the speed
of orthogonality doesn't affected as one increases $\eta $ (See Fig.(7b),
where we set $\eta =0.8$).

\section{Conclusion}

The dynamics of a charged qubit interacts with a cavity mode prepared
initially in either Fock state or Binomial states is investigated. The
computational speed is studied by means of the orthogonality speed. The
effect of the field and the charged qubit parameters is investigated. We
show that, the detuning parameter and the number of photons inside the
cavity play essential roles on controlling the speed of orthogonality and
consequently the computational speed. However, larger values of the detuning
and the number of photons, lead to increase the number of orthogonality and
decrease the orthogonality time and consequently decrease the computational
time, i.e., the speed of computations is increased. On the other hand, the
effect of the the coupling between the charged qubit and the cavity mode is
different. It is shown that, as one increases the coupling parameter, the
number of orthogonality is increased and the orthogonality time is
increased, i.e., the speed of computation is decreased.

For binomial case, the parameter $\eta $, has almost no effect on the speed
of orthogonality, while the coupling constant between the field and the
Cooper pair has a noticeable effect. For a large value of $g$, the number of
orthogonality is decreased and consequently the computational speed is
decreased. \newline

\textbf{Acknowledgement}: we are grateful for the helpful comments given by
the referees.

\end{document}